\documentclass[conference]{IEEEtran}
\IEEEoverridecommandlockouts

\usepackage{cite}
\usepackage{amsmath,amssymb,amsfonts}
\usepackage{algorithmic}
\usepackage{graphicx}
\usepackage{textcomp}
\usepackage{xcolor}
\usepackage{multirow}
\usepackage{tabularx} 
\usepackage{booktabs} 
\usepackage{makecell}
\usepackage{soul}
\def\BibTeX{{\rm B\kern-.05em{\sc i\kern-.025em b}\kern-.08em
    T\kern-.1667em\lower.7ex\hbox{E}\kern-.125emX}}
\begin{document}

\title{Audio-based Anomaly Detection in Industrial Machines Using Deep One-Class Support Vector Data Description}

\author{\IEEEauthorblockN{1\textsuperscript{st} Sertac Kilickaya}
\IEEEauthorblockA{\textit{Dept. of Computing Sciences}\\
\textit{Tampere University}\\
Tampere, Finland \\
sertac.kilickaya@tuni.fi}
\and
\IEEEauthorblockN{2\textsuperscript{nd} Mete Ahishali}
\IEEEauthorblockA{\textit{Dept. of Computing Sciences}\\
\textit{Tampere University}\\
Tampere, Finland \\
mete.ahishali@tuni.fi}
\and
\IEEEauthorblockN{3\textsuperscript{rd} Cansu Celebioglu}
\IEEEauthorblockA{\textit{Dept. of Information Engineering} \\
\textit{University of Padova}\\
Padova, Italy \\
cansu.celebioglu@studenti.unipd.it}
\and
\IEEEauthorblockN{4\textsuperscript{th} Fahad Sohrab}
\IEEEauthorblockA{\textit{Dept. of Computing Sciences}\\
\textit{Tampere University}\\
Tampere, Finland \\
fahad.sohrab@tuni.fi}
\and
\IEEEauthorblockN{5\textsuperscript{th} Levent Eren}
\IEEEauthorblockA{\textit{Dept. of Electrical } \\
\textit{and Electronics Engineering}\\
\textit{Izmir University of Economics}\\
Izmir, Turkey \\
levent.eren@ieu.edu.tr}
\and
\IEEEauthorblockN{6\textsuperscript{th} Turker Ince}
\IEEEauthorblockA{\textit{Dept. of Media Engineering }\\
\textit{and Technology}\\
\textit{German International University}\\
Berlin, Germany \\
turker.ince@giu-berlin.de}
\and
\IEEEauthorblockN{7\textsuperscript{th} Murat Askar}
\IEEEauthorblockA{\textit{Dept. of Electrical } \\
\textit{and Electronics Engineering}\\
\textit{Izmir University of Economics}\\
Izmir, Turkey \\
murat.askar@ieu.edu.tr}
\and
\IEEEauthorblockN{8\textsuperscript{th} Moncef Gabbouj}
\IEEEauthorblockA{\textit{Dept. of Computing Sciences}\\
\textit{Tampere University}\\
Tampere, Finland \\
moncef.gabbouj@tuni.fi}
}

\maketitle

\begin{abstract}
The frequent breakdowns and malfunctions of industrial equipment have driven increasing interest in utilizing cost-effective and easy-to-deploy sensors, such as microphones, for effective condition monitoring of machinery. Microphones offer a low-cost alternative to widely used condition monitoring sensors with their high bandwidth and capability to detect subtle anomalies that other sensors might have less sensitivity. In this study, we investigate malfunctioning industrial machines to evaluate and compare anomaly detection performance across different machine types and fault conditions. Log-Mel spectrograms of machinery sound are used as input, and the performance is evaluated using the area under the curve (AUC) score for two different methods: baseline dense autoencoder (AE) and one-class deep Support Vector Data Description (deep SVDD) with different subspace dimensions. Our results over the MIMII sound dataset demonstrate that the deep SVDD method with a subspace dimension of 2 provides superior anomaly detection performance, achieving average AUC scores of 0.84, 0.80, and 0.69 for 6 dB, 0 dB, and -6 dB signal-to-noise ratios (SNRs), respectively, compared to 0.82, 0.72, and 0.64 for the baseline model. Moreover, deep SVDD requires 7.4 times fewer trainable parameters than the baseline dense AE, emphasizing its advantage in both effectiveness and computational efficiency.
\end{abstract}


\section{Introduction}
In today’s fast-changing industrial environment, ensuring the reliability and efficiency of machinery is crucial to maintain uninterrupted operations, optimize productivity, and minimize direct and indirect financial losses. Unplanned machinery failures not only cause costly downtimes, but can also lead to extensive damage to equipment and potential safety hazards to personnel. With the increasing complexity and automation of industrial systems, the demand for comprehensive and cost-effective condition monitoring solutions has increased dramatically.

\begin{figure*}[htbp]
\centerline{\includegraphics[width=1.0\linewidth]{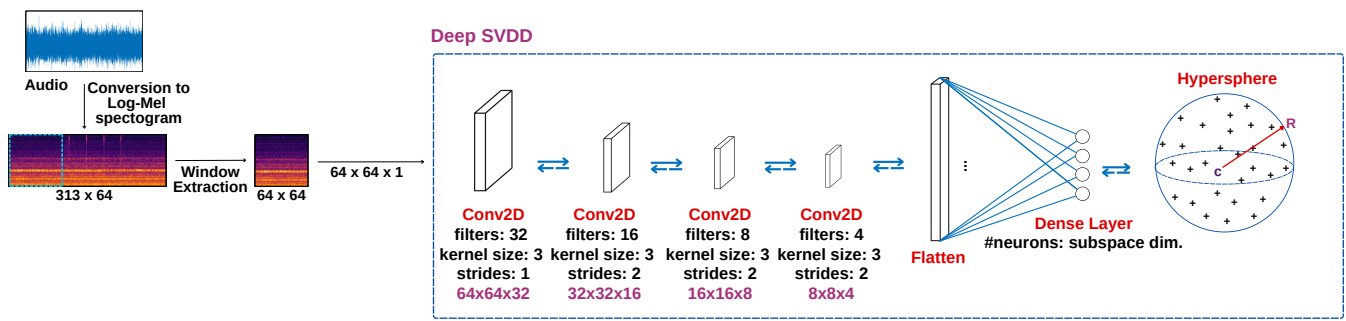}}
\caption{Deep SVDD model is illustrated with its preprocessing steps.}
\label{fig1}
\end{figure*}

Recently, machine audio-based condition monitoring has gained popularity due to its numerous advantages \cite{bib1}. Acquiring audio data is non-intrusive and does not require direct contact with the machine. Furthermore, audio sensors are more affordable and their installations are easier compared to vibration sensors. Naturally, audio-based anomaly detection is a significantly cost-effective solution, especially for large-scale industrial applications where numerous sensors may be required. Thanks to recent advances in machine learning and audio processing, the accuracy and reliability of sound-based fault detection have been substantially improved, making it a compelling alternative to traditional methods \cite{bib2, bib3, bib4, bib5, bib6}.

Acoustic anomaly detection (AAD) aims to identify unusual patterns in audio data. In a one-class classification (OCC) problem, samples from a single class are used for training. Thus, when applied to AAD, it can be viewed as a OCC task, as the training process relies solely on normal samples. OCC tasks are typically addressed through three main approaches: density estimation, reconstruction, and boundary description \cite{bib7}. Among these, the boundary description approach is particularly prevalent. The boundary description approach seeks to establish a boundary around the target data. Key techniques include Support Vector Data Description (SVDD) \cite{bib8}, which encloses the target class within a hypersphere with its extensions using different kernel methods, and One-Class Support Vector Machine (OC-SVM) \cite{bib9}, which defines a hyperplane to maximize the margin between the data and the origin. Next, Subspace Support Vector Data Description (S-SVDD) \cite{bib7} has been proposed for joint subspace learning and classification. Accordingly, S-SVDD utilizes subspace learning to project data into a lower-dimensional feature space that is optimized for the OCC tasks. As a result, S-SVDD eliminates the need for separate dimensionality reduction or feature selection procedures within the classification framework, and it has demonstrated superior anomaly detection performance compared to SVDD in various studies \cite{bib10, bib11, bib12}. Furthermore, deep Support Vector Data Description (deep SVDD) is an extension of traditional SVDD that leverages deep neural networks to model more complex decision boundaries \cite{bib13}. In deep SVDD, a deep neural network maps input data into a lower-dimensional feature space, where it learns a hypersphere around the features of normal data points. This hypersphere is then optimized to minimize its volume while encompassing the majority of target data. Several studies used deep SVDD and its variants for anomaly detection in industrial machines as well \cite{bib14, bib15}.

In this study, we compare two different deep learning-based anomaly detection methods over the MIMII sound dataset \cite{bib16}, targeting malfunctioning industrial machines. Log-Mel spectrograms of machinery sounds are used as input to evaluate the performance of the following approaches: a baseline dense AE \cite{bib16} and one-class deep SVDD with varied number of subspace dimensions. Anomaly detection performance is measured using the area under the curve (AUC) score, and further comparisons are performed regarding computational efficiency and complexity.

\section{Dataset and Feature Extraction}
MIMII sound dataset was captured using a TAMAGO-03 circular microphone array from System In Frontier Inc. \cite{bib16}. The dataset includes recordings from four types of machines: pumps, fans, valves, and slide rails. For each machine type, recordings were collected from various product models that are specified by model ID labels. The microphone array was positioned 50 cm away from the slide rail, fan, and pump, and 10 cm from the valve. Each audio segment is 10 seconds long, with 8 channels. They were recorded separately as 16-bit audio at a 16 kHz sampling rate in a reverberant environment while the machines were running. Real-life anomalies such as contamination, leakage, rotating unbalance, and rail damage were introduced to the machines. The background noise from various real-world factory settings was combined with the target machine sounds to simulate noisy environments. Generated noise-mixed audio data have three different SNR values of $-6$ dB, $0$ dB, and $6$ dB, and they were created by adding the adjusted background noise to the machine audio segments. As a result, three distinct datasets were generated for varied SNRs values, each containing the number of samples specified in \cite{bib16}.

\begin{figure}[htbp]
\centerline{\includegraphics[width=1.0\linewidth]{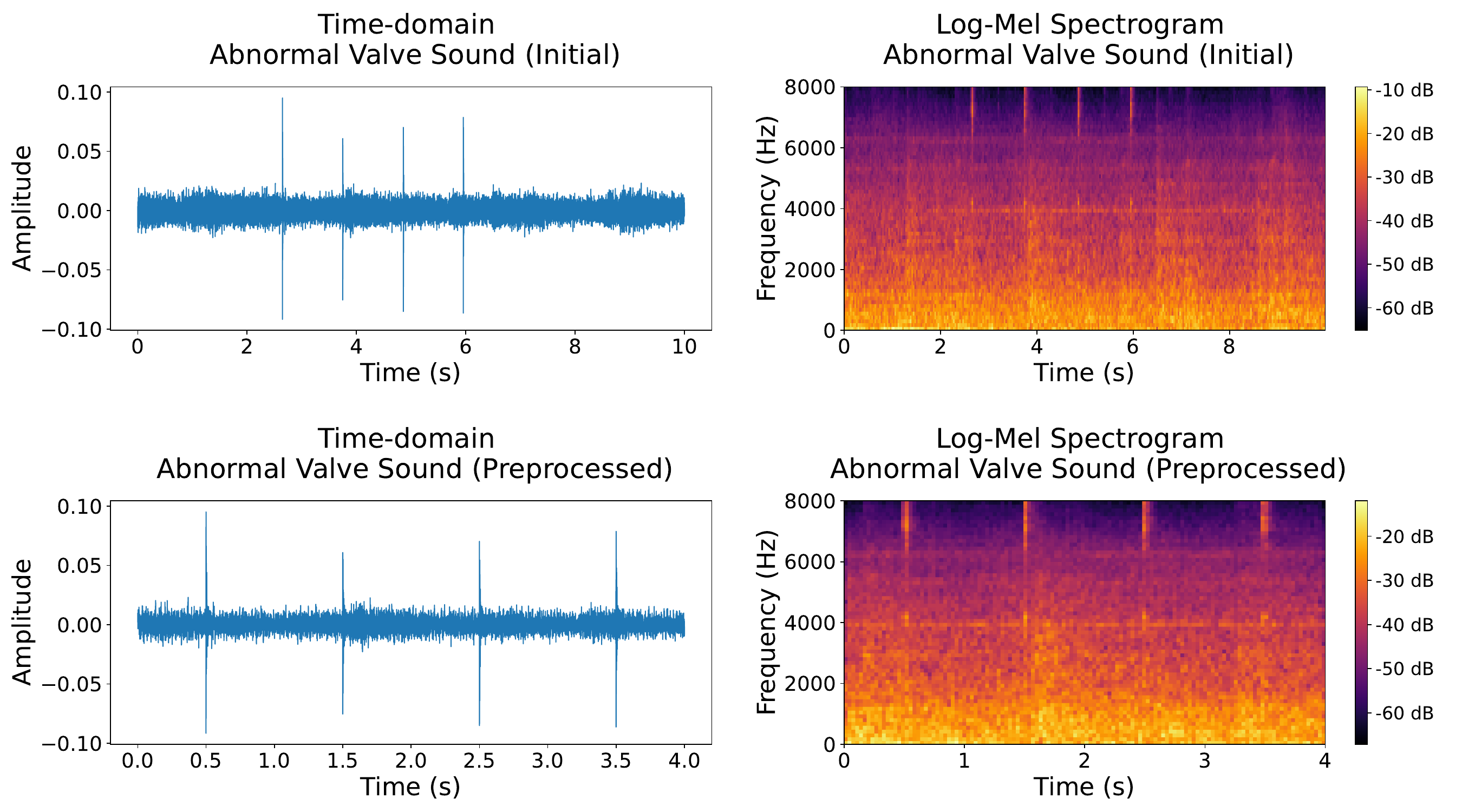}}
\caption{Examples of initial and preprocessed time-domain waveforms with corresponding log-Mel spectrograms for valve data.}
\label{fig2}
\end{figure}

For feature extraction, we used the audio data from the first microphone channel positioned at 0 degrees, similar to the baseline approach \cite{bib16}. Prior to the feature extraction process, we addressed the non-stationary nature of valve data by removing silent sections. For every 10-second valve sound segment, we perform peak detection, isolate 1-second segments around the detected peaks, and then overwrite the sample with the processed version. For other machine types, these steps are omitted. Fig. \ref{fig2} shows examples of both initial and processed time-domain waveforms using the proposed preprocessing procedure, and their corresponding log-Mel spectrograms for the valve data. Log-Mel spectrograms, widely used for capturing the time-frequency representation of audio signals, were generated with a frame size of $1024$ samples, a hop size of $512$ samples, and a filter bank of $64$ Mel filters, producing log-Mel spectrograms of size $313\times64$. To prepare the features for the deep learning models used in this study, we performed additional preprocessing on the log-Mel spectrogram. For the baseline dense AE, five consecutive frames were concatenated to form a 320-dimensional input feature vector \cite{bib16}. However, for the deep SVDD models, the log-Mel spectrogram is directly given to the model as 2D plane and it is not flattened. Instead, the resulting log-Mel spectrograms are divided into $64\times64$ windows without overlap, where each window represents approximately two seconds of audio. Zero padding is applied to the last cropped spectrogram to ensure that all windows conform to the $64\times64$ dimension. Using a shorter window length allows for the extraction of more uniform segments of the log-Mel spectrogram, and it enables the system to detect anomalies with just two seconds of recorded audio, which is crucial for real-time anomaly detection applications.

\section{Deep Support Vector Data Description}
In this section, we provide an overview of the theoretical foundation of the deep SVDD method. In SVDD, the algorithm aims to find the smallest enclosing hyperball that contains only the target (normal) observations in a high-dimensional space \cite{bib8}. Given the target data points in the high-dimensional feature space $\mathbb{R}^D$, the goal is to determine the center $\mathbf{c} \in \mathbb{R}^D$ and the radius $R$ of the hypersphere by minimizing the following objective:
\begin{equation} 
F(R, \mathbf{c})=R^{2}+C\sum_{i=1}^{N}\xi_{i},
\end{equation}
such that the following conditions are met, i.e.:
\begin{equation}
\Vert \mathbf{x}_i - \mathbf{c} \Vert_2^2 \leq R^2 + \xi_i, \ \xi_i \geq 0, \ i = 1, \ldots, N.
\end{equation}
The parameter $C > 0$ acts as a regularization factor that balances the trade-off between the volume of the hypersphere and the training error caused by allowing outliers in the class description, and $\xi_i$ denotes the set of slack variables.

To address the problem of overfitting in high-dimensional feature space, Subspace Support Vector Data Description (S-SVDD) was proposed in \cite{bib7}, and it aims to identify a lower-dimensional feature space ($d < D$) that optimally represents the target class, by multiplying the input features with a matrix $ \mathbf{Q}\in\mathbb{R}^{d \times D} $ as follows:
\begin{equation}
\mathbf{y}_i = \mathbf{Qx}_i, \ \text{s.t.} \ \Vert \mathbf{Qx}_i - \mathbf{a} \Vert_2^2 \leq R^2 + \xi_i, \ \xi_i \geq 0, \ i = 1, \ldots, N.
\end{equation}

In S-SVDD, the objective remains to find a minimal hypersphere that encloses the target observations, but this time in the reduced feature space $d$. The modification to the original SVDD formulation is the inclusion of the matrix $\mathbf{Q} $. Lagrange-based optimization techniques are employed to determine the matrix $ \mathbf{Q} $, with $ \mathbf{Q} $ being updated iteratively using a gradient descent approach \cite{bib7}.

Similar to S-SVDD, deep SVDD reduces the dimensionality of high-dimensional features by training a neural network to map them into a lower-dimensional space \cite{bib13}. While both methods aim to reduce the feature space, S-SVDD does so without the use of a neural network, distinguishing it from deep SVDD's neural-based approach. In deep SVDD, the network outputs are then constrained within a hypersphere, with the goal of minimizing its volume to encapsulate the most important features of the target data. Two versions of deep SVDD were introduced in the original paper \cite{bib13}, and \textit{soft-boundary} deep SVDD objective can be written as:
\begin{equation}
\begin{aligned}
\min_{R, \mathbf{W}} \quad & R^2 + C \sum_{i=1}^N \max \left\{ 0, \left\| \phi(\mathbf{x}_i; \mathbf{W}) - \mathbf{c} \right\|^2 - R^2 \right\} \\
& + \frac{\lambda}{2} \| \mathbf{W} \|_F^2,
\end{aligned}
\end{equation}
where $\phi$ represents the feature mapping applied to the input $\mathbf{x}_i$ with neural network parameters $\mathbf{W}$. Minimizing \( R^2 \) in \textit{soft-boundary} deep SVDD reduces the hypersphere's volume, while the second term penalizes points outside the sphere. The hyperparameter $C$ controls the trade-off between sphere volume and boundary violations. The final term acts as a weight decay regularizer for the network parameters \( \mathbf{W} \), with \( \lambda > 0 \) as the hyperparameter, and \( \| \cdot \|_F \) denotes the Frobenius norm.

\begin{table*}[]
\caption{Average AUCs across 3 different runs for each method on the MIMII dataset}
\footnotesize\setlength{\tabcolsep}{8.0pt}
\resizebox{\textwidth}{!}{%
\begin{tabular}{@{}cccccccccccccccc@{}}
\toprule
\textbf{Machine Type} &
  \multicolumn{3}{c}{\textbf{Valve}} &
  \multicolumn{3}{c}{\textbf{Pump}} &
  \multicolumn{3}{c}{\textbf{Fan}} &
  \multicolumn{3}{c}{\textbf{Slide Rail}} &
  \multicolumn{3}{c}{\textbf{All Machines AVG.}} \\ \midrule
\textbf{Input SNR} &
  \textbf{6dB} &
  \textbf{0dB} &
  \textbf{-6dB} &
  \textbf{6dB} &
  \textbf{0dB} &
  \textbf{-6dB} &
  \textbf{6dB} &
  \textbf{0dB} &
  \textbf{-6dB} &
  \textbf{6dB} &
  \textbf{0dB} &
  \textbf{-6dB} &
  \textbf{6dB} &
  \textbf{0dB} &
  \textbf{-6dB} \\ \midrule
\begin{tabular}[c]{@{}c@{}}Baseline Dense AE\\ \cite{bib16}\end{tabular} &
  0.670 &
  0.613 &
  0.555 &
  0.805 &
  0.705 &
  0.660 &
  0.913 &
  0.795 &
  0.663 &
  \textbf{0.878} &
  0.780 &
  0.700 &
  0.816 &
  0.723 &
  0.644 \\ \midrule
\begin{tabular}[c]{@{}c@{}}Baseline Dense AE\\ (Preprocessed Valve)\end{tabular} &
  0.673 &
  0.626 &
  0.589 &
  0.811 &
  0.719 &
  0.647 &
  0.916 &
  0.802 &
  \textbf{0.675} &
  0.872 &
  \textbf{0.795} &
  0.705 &
  0.818 &
  0.736 &
  0.654 \\ \midrule
\begin{tabular}[c]{@{}c@{}}One-Class Deep SVDD\\ (subspace dim: 2)\end{tabular} &
  0.772 &
  0.786 &
  \textbf{0.742} &
  0.802 &
  \textbf{0.781} &
  0.695 &
  \textbf{0.936} &
  \textbf{0.834} &
  0.647 &
  0.867 &
  0.791 &
  0.670 &
  \textbf{0.844} &
  \textbf{0.798} &
  0.689 \\ \midrule
\begin{tabular}[c]{@{}c@{}}One-Class Deep SVDD\\ (subspace dim: 4)\end{tabular} &
  0.799 &
  0.774 &
  0.713 &
  0.782 &
  0.767 &
  0.702 &
  0.932 &
  0.809 &
  0.656 &
  0.864 &
  0.789 &
  \textbf{0.718} &
  \textbf{0.844} &
  0.785 &
  \textbf{0.697} \\ \midrule
\begin{tabular}[c]{@{}c@{}}One-Class Deep SVDD\\ (subspace dim: 8)\end{tabular} &
  \textbf{0.804} &
  \textbf{0.789} &
  0.712 &
  \textbf{0.837} &
  0.750 &
  \textbf{0.706} &
  0.904 &
  0.823 &
  0.674 &
  0.829 &
  0.770 &
  0.664 &
  0.843 &
  0.783 &
  0.689 \\ \bottomrule
\end{tabular}%
}
\label{tab1}
\end{table*}

Since the majority of the training data is typically normal, \textit{one-class} deep SVDD tries to minimize a simpler objective function during training as follows:
\begin{equation}
\min_{\mathbf{W}} \left\{ \frac{1}{N} \sum_{i=1}^N \left\| \phi(\mathbf{x}_i; \mathbf{W}) - \mathbf{c} \right\|^2 + \frac{\lambda}{2} \| \mathbf{W} \|_F^2 \right\}.
\end{equation}
Basically, \textit{one-class} deep SVDD tightens the sphere by minimizing the average distance of all data representations from the center $\mathbf{c}$. It uses a quadratic loss to penalize the distance of each network representation \( \phi(\mathbf{x}_i; \mathbf{W}) \) from the center.

For a test point \( \mathbf{x} \), the anomaly score \( s \) for both deep SVDD variants is defined by the distance to the hypersphere center: 
\begin{equation}
s(\mathbf{x}) = \left\| \phi(\mathbf{x}; \mathbf{W}^*) - \mathbf{c} \right\|^2,
\label{eq16}
\end{equation}
where \( \mathbf{W}^* \) are the trained network parameters.

\section{Experiments and Results}
We compare the performance of the baseline dense AE and the \textit{one-class} deep SVDD on the MIMII dataset, adhering to the training procedures outlined in \cite{bib16}. For each machine type, model ID and input SNR, we partitioned the audio data into training and test datasets. All anomalous segments were assigned to the test dataset, together with an equal number of randomly selected normal segments. Only the remaining normal segments were used for training without any anomalous segments. During training, 10\% of the training data was used for validation. Training was stopped if the validation loss did not improve for 10 consecutive epochs, and the model with the lowest validation loss was used for testing. In the baseline dense AE \cite{bib16}, the input of the model is a concatenated sequence of 5 consecutive frames from the log-Mel spectrogram. With 64 Mel filter banks, the resulting input is a 320-dimensional feature vector. Using the same training configuration in \cite{bib16}, we also trained the dense AE with the preprocessed valve data, and AUC results are presented in Table III under the label ``Baseline Dense AE (Valve Preprocessed)". In contrast, the ``Baseline Dense AE" column shows the AUCs obtained using the original unprocessed data.

On the other hand, the one-class deep SVDD model, shown in Fig. \ref{fig1}, processes $64\times64$ log-Mel spectrograms. For deep SVDD, we exclude bias terms from all network units to avoid hypersphere collapse \cite{bib13}, and all layers use the LeakyReLU activation function with negative slope coefficient of 0.2. After 2D convolutional layers, the extracted features are flattaned and passed through a dense layer. To assess the impact of subspace dimension, we configure the dense layer with 2, 4, and 8 neurons in three different deep SVDD models. We set the hypersphere center \( c \) to the mean of the mapped data following an initial forward pass, as suggested by \cite{bib13}. The weight decay hyperparameter was set to $\lambda = 10^{-5}$. Before training, the log-Mel spectrograms were normalized by subtracting the mean of the training set and scaling by its standard deviation. All deep SVDD models were trained for a maximum of 50 epochs, and the weights were updated using the Adam optimizer with a learning rate of 0.0005. The batch size is 32.

During testing, anomaly detection for each 10-second sound segment was based on two metrics: reconstruction error for the dense AE and distance from the hypersphere center for the one-class deep SVDD models. These metrics are averaged over the 10-second duration for each sample. AUC scores were then computed on the test dataset by evaluating the performance of the models at all possible thresholds.

The average AUC scores across all model IDs are presented in Table \ref{tab1}. The one-class deep SVDD model (with a subspace dimension of 2) achieved average AUCs of 0.84, 0.80, and 0.69 for SNRs of 6 dB, 0 dB, and -6 dB, respectively, averaged across all machine types. These results surpass the baseline dense AE model with AUCs of 0.82, 0.72, and 0.64 for the same SNR levels. We also observe a notable performance improvement when the valve data is preprocessed. While preprocessing the valve data enhances the AUC scores for the baseline dense AE, it cannot reach the anomaly detection performance of the deep SVDD models. The subspace dimension for deep SVDD introduces only minimal variance in the results, demonstrating the model's robust performance. Table \ref{tab2} summarizes the number of trainable parameters and inference duration for each method evaluated, where durations have been averaged over 100 different runs on an NVIDIA RTX 3070 GPU. It included the time for calculating reconstruction error for the AE and distances for deep SVDD. Deep SVDD models have fewer trainable parameters, but they require slightly more time for inference compared to the baseline dense AE.

\begin{table}[htbp]
\caption{Number of trainable parameters and inference duration for each method}
\resizebox{\columnwidth}{!}{%
\begin{tabular}{@{}ccc@{}}
\toprule
\textbf{Model}             & \textbf{Trainable Params} & \textbf{Inference Time (ms)} \\ \midrule
Baseline Dense AE \cite{bib16} & 50,760                    & 2.563                        \\
One-Class Deep SVDD (dim: 2)          & 6,848                     & 2.644                        \\
One-Class Deep SVDD (dim: 4)          & 7,360                     & 2.694                        \\
One-Class Deep SVDD (dim: 8)          & 8,384                     & 2.702                        \\ \bottomrule
\end{tabular}%
}
\label{tab2}
\end{table}

\section{Conclusion}
The growing interest in using microphones for monitoring industrial machinery stems from their high sensitivity, affordability, and straightforward deployment. This approach mirrors the common practice of human operators identifying machinery anomalies by sound, highlighting the natural alignment between acoustic monitoring and human perceptual skills. To contribute to this field, this study compares two different methods: baseline dense AE and one-class deep SVDD using the MIMII sound dataset. Our results demonstrate that deep SVDD models outperform the baseline dense AE in terms of AUC scores. Specifically, the deep SVDD model with a subspace dimension of 2 achieved the highest average AUCs across all machines. Furthermore, the proposed preprocessing procedure for valve data has further improved the results. Overall, deep SVDD offers a superior, cost-effective solution for acoustic anomaly detection, advancing monitoring solutions for industrial machines. For future work, we plan to explore multimodal and graph-embedded approaches for deep SVDD, as described in \cite{bib17, bib18}.

\section*{Acknowledgment}
This work was supported by the NSF-Business Finland project AMALIA and H2TRAIN research program under the Horizon Europe Framework.

\bibliography{refs}
\bibliographystyle{IEEEtran}

\end{document}